# A Quinary Coding and Matrix Structure-based Channel Hopping Algorithm for Blind Rendezvous in Cognitive Radio Networks

Qinglin Liu[1], Zhiyong Lin[1]*, Zongheng Wei[1], Jianfeng Wen[1], Congming Yi[1], Hai Liu[2]
[1]School of Computer Science, Guangdong Polytechnic Normal University, Guangzhou, China
Email: lql6315@163.com, zhiyong.lin@qq.com, {zhwei, wenjf}@gpnu.edu.cn, cm_yi@foxmail.com
[2]Department of Computing, The Hang Seng University of Hong Kong, Hong Kong, China
Email: hliu@hsu.edu.hk
*Corresponding Author: Zhiyong Lin, Email: zhiyong.lin@qq. com

*Abstract*—The multi-channel blind rendezvous problem in distributed cognitive radio networks (DCRNs) refers to how users in the network can hop to the same channel at the same time slot without any prior knowledge (i.e., each user is unaware of other users' information). The channel hopping (CH) technique is a typical solution to this blind rendezvous problem. In this paper, we propose a quinary coding and matrix structure-based CH algorithm called QCMS-CH. The QCMS-CH algorithm can guarantee the rendezvous of users using only one cognitive radio in the scenario of the asynchronous clock (i.e., arbitrary time drift between the users), heterogeneous channels (i.e., the available channel sets of users are distinct), and symmetric role (i.e., all users play a same role). The QCMS-CH algorithm first represents a randomly selected channel (denoted by R) as a fixed-length quaternary number. Then it encodes the quaternary number into a quinary bootstrapping sequence according to a carefully designed quaternary-quinary coding table with the prefix "R00". Finally, it builds a CH matrix column by column according to the bootstrapping sequence and six different types of elaborately generated subsequences. The user can access the CH matrix row by row and accordingly perform its channel hopping to attempt to rendezvous with other users. We prove the correctness of QCMS-CH and derive an upper bound on its Maximum Time-to-Rendezvous (MTTR). Simulation results show that the QCMS-CH algorithm outperforms the state-of-the-art in terms of the MTTR and the Expected Time-to-Rendezvous (ETTR).

## I. INTRODUCTION

IN recent years, due to the rapidly growth of wireless devices, the problem of spectrum scarcity has become more and more serious. Existing studies have shown that many licensed spectrum bands are underutilized [1]. Therefore, cognitive radio technique came into being, which can make more efficient and better use of the precious spectrum band. In distributed cognitive radio networks (DCRNs), there are a set of channels with different frequencies and two types of spectrum users, i.e., primary users (PUs) and secondary users (SUs). PUs have the priority to use the licensed channels, and SUs can only utilize those channels that are not currently occupied by PUs [2]. At any moment, the channels which can be used by SUs are called available channels. If two or more SUs in a proximity need to communicate with each other, they must hop to and access the same available channel at the same time to set up a communication link. In this paper, we study the problem of how to set up an initial communication link between SUs in DCRNs.

For convenience, we refer to SUs as users. In addition, we assume that time is divided into slots of fixed-length. In DCRNs, if the users are unaware of each other's available channels and local clocks, and have no pre-assigned channel hopping (CH) strategies, it is hard to guarantee the users to hop to the same channel at the same time slot (i.e., set up a communication link). This challenging problem is referred to as the multi-channel blind rendezvous problem (or the rendezvous problem for short) [3], [4]. A number of works have been proposed on how to design CH sequences for blind rendezvous in DCRNs. The desirable CH algorithms should require as few prerequisites as possible. For example, it is not preferable to assume time-synchronization of network, multiple radios at each SU, different roles of SUs, and even availability of user IDs. Meanwhile, the CH algorithms are expected to give good performance in terms of the maximum time-to-rendezvous (MTTR) (i.e., the users are guaranteed to achieve rendezvous in a bounded time in the worst case), as well as the average time-to-rendezvous (i.e., Expected TTR (ETTR)).

In this paper, we propose a novel approach of generating CH sequences, named QCMS-CH (short for Quinary Coding and Matrix Structure based Channel Hopping), to solve the rendezvous problem. The CH sequence generated by the QCMS-CH algorithm can be represented by a matrix (CH matrix). Specifically, in QCMS-CH, each user first randomly selects an available channel and transforms its index (denoted by *R*) into a fixed-length quaternary number, then uses a carefully designed quaternary-quinary coding table to convert the quaternary number into a quinary number. Then, the user adds the prefix "R00" in front of the quinary number to construct a bootstrapping sequence. The length of this bootstrapping sequence determines the number of columns of the CH matrix, and each element of the bootstrapping sequence, which may be one of {R, 0, 1, 2, 3, 4}, corresponds to a column of the CH matrix. Next, the CH matrix is built column by column. Since there are only six different types of columns, i.e., R-type and $\lambda$-type ($\lambda \in \{0, 1, ..., 4\}$, six types of subsequences associated with R-type and $\lambda$-type can be elaborately constructed based on the available channels of the user. With the construction of subsequences, the CH matrix can be eventually built by repeating the corresponding subsequence in each column, e.g., a 0-type subsequence is repeated in a column associated with 0, and a 1-type subsequence is repeated in a column associated with 1, etc. Once the CH matrix is built, the user can access the matrix row by row and accordingly perform its channel hopping to attempt rendezvous with other users.

The main contributions of this paper are summarized as follows:

1) The proposed QCMS-CH algorithm solves the blind rendezvous problem in general scenario because it is applicable to single-radio at each user, asynchronous clocks, heterogeneous available channels, and symmetric roles among the users. Moreover, QCMS-CH depends on only the available channels of user and does not need using user ID.

2) To the best of our knowledge, compared with existing similar algorithms [5]-[9] which construct the bootstrapping sequences based on the binary representation of the randomly selected channel, QCMS-CH is the first one that directly uses the quaternary representation of the channel and further adopts the efficient quaternary-quinary coding to generate the bootstrapping sequence with shorter length. Given the number of potential available channels $N$, the bootstrapping sequence generated by QCMS-CH has length of $L = 2 \times \lceil \lceil \log_4 N \rceil /2 \rceil + 3$, which does not exceed 9 when $N \leq 1024$. Thanks to the shorter bootstrapping sequence and the elaborately constructed different types of subsequences, QCMS-CH provides very competitive theoretical performance with the MTTR upper-bounded by $\max\{(P_A+4)(P_B+6), (P_B+4)(P_A+6)\} \times L$, where $P_A$ and $P_B$ are the smallest prime number not less than the number of available channels of user $A$ and user $B$, respectively.

3) We conduct extensive simulations to verify the effectiveness of the QCMS-CH algorithm. Compared with several most recent representative algorithms, QCMS-CH performs better in terms of both MTTR and ETTR. According to the simulation results, QCMS-CH can save more than 5% of the average TTR than the state-of-the-art in many cases.

The rest of this paper is organized as follows. In Section II, we review related works. In Section III, we propose the system model and the formulation of the problem. In Section IV, we describe the QCMS-CH algorithm in details. In Section V, we present extensive simulation results. We conclude this paper in the Section VI.

## II. RELATED WORKS

In cognitive radio networks, rendezvous is a prerequisite for communication between users. The traditional solution is to use the Common Control Channel (CCC) to achieve rendezvous. However, as the number of users increases, the CCC scheme suffers from congestion. In addition, CCC is vulnerable to external interference and malicious attacks [10]. Therefore, the rendezvous algorithms without CCC have attracted extensive attention of researchers. In this section, we briefly review some representative blind rendezvous algorithms, which are divided into different classes based on their assumptions.

*Symmetric role/Asymmetric role*: This classification is based on whether users can be divided into different roles. In the asymmetric algorithm (e.g. [11], [12] and [13]), the users are divided into senders and receivers, and generate different CH sequences according to different roles. In the symmetric algorithm (e.g. [14] and [15]), the users are all in the same role, so their CH sequences are generated according to the same strategy. In DCRNs, all users are equal and no roles are preassigned to each user before they complete the rendezvous. Therefore, the CH algorithm with symmetric roles is more widely used in DCRNs.

*Homogeneous channels/Heterogeneous channels*: If the available channel sets of two users are exactly the same, it is called a homogeneous model; otherwise, it is called a heterogeneous model. In reality, if two users are geographically close, they are likely to have the same set of available channels; if two users are far apart, they tend to have different sets of available channels. The early works, say [16] and [14], consider both of these two models. However, since the homogeneous mode can be viewed as a special case of the heterogeneous model, and the latter is more practical, most of recent research has placed focus on the heterogeneous model.

*Synchronous clock/Asynchronous clock*: Synchronous clock algorithms (e.g., [17] and [18]) assume that all users can synchronize their time through a common time source (e.g., GPS), so their CH sequences can be started at the same time. However, in DCRNs, there may have no common time source for reference due to the hardware limitation, and it would be hard for the users to synchronize in many cases. Therefore, asynchronous algorithms (e.g., [6], [7] and [14]) are more general and practical in DCRNs.

*Onymous ID/Anonymous ID*: The onymous algorithm (e.g., [19], [20] and [21]) utilizes the user's unique identifier (ID) to generate their CH sequences. However, once the user's ID is exposed, the user would be vulnerable to external malicious attacks. Therefore, in the consideration of security, the anonymous algorithms without using user ID (e.g., [4], [6] and [14]) are more attractive than the onymous algorithms.

*Single-radio/Multi-radio*: A user can be equipped with multiple radio transceivers, so they can hop to multiple channels simultaneously at a time slot. The multi-radio algorithms (e.g., [5], [22] and [23]) can greatly reduce the time-to-rendezvous (TTR), meanwhile bearing the increase of hardware cost. From the perspective of economic feasibility, the single-radio algorithms (e.g., [8], [9] and [14]) are more preferred than the multi-radio algorithms. However, it is also more challenging to design efficient rendezvous algorithms for users with single radio.

*Global channel set/Local channel set*: The early works, e.g., [14] and [16], generate the CH sequences based on the global channel set, i.e., the whole set of all potentially available channels. However, in reality, the channels actually available to a user usually take up a small portion of the whole channel set. The CH sequences generated based on the whole channel set may consist of many unavailable channels, and thus usually have much large length and TTR. More recently, researchers have pay increasing intention to the design of CH sequences based on the local channel set (i.e., only the channels available to each user) for better efficiency [5]-[9], [24], although this direction is more challenging than that based on the global channel set.

In this paper, we mainly study the construction of CH sequences for efficient blind rendezvous of users in DCRNs under a general scenario by consideration of symmetric role, heterogeneous channels, asynchronous clock, anonymous ID, single-radio, and local channel set.

## III. SYSTEM MODEL AND PROBLEM FORMULATION

### A. System Model

We assume that a DCRN consists of $N$ non-overlapping channels with universal indices from 1 to $N$ and $M$ ($M \geq 2$) users. Each user is equipped with one cognitive radio transceiver, with which the user can sense the state of any channel and switch between the perceived idle channels (i.e., its available channels). We use $C=\{1, 2, ..., N\}$ and $U$ to denote the set of channels (or more precisely, channel indices) and the set of users, respectively. We use $C_A$ ($C_A \subseteq C$, $C_A \neq \phi$) to denote the set of channels available to user $A$ ($A \in U$), and $|C_A|$ is the cardinality of $C_A$. We assume that there is at least one common available channel between any two users, that is, for $\forall A, B \in U$, $G_{AB}=|C_A \cap C_B| \geq 1$.

The network is time-slotted and each slot has a length of $2\Delta$, where $\Delta$ represents the duration required for message exchange and link establishment between two users when they rendezvous on a same channel. The value of $\Delta$ is usually set as 10 ms [14]. The reason for doubling the slot length is that, when the slots of two users are not aligned, it can still guarantee that there is at least $\Delta$ time overlap for accomplishing link establishment when the two users rendezvous on a same channel. Via such setting, the slots of any pair of users can be regarded as aligned for the design of CH sequences [14].

### B. Problem Formulation

As most of existing relevant works, we focus on the two-user rendezvous problem. The rendezvous problem with more than two users can be solved by some extension of two-user rendezvous algorithms or specific design of multi-user rendezvous algorithm [14], [25], which is beyond the scope of this paper. Let $S_A = \{S_A^1, S_A^2, ..., S_A^{T_A}\}$ and $S_B = \{S_B^1, S_B^2, ..., S_B^{T_B}\}$ denote the CH sequences of user $A$ and user $B$, respectively, where $T_A$ and $T_B$ denote the period of the CH sequences of user $A$ and user $B$, respectively. In the asynchronous clock scenario, we suppose user $A$ starts its channel hopping with $\delta$ ($\delta \in N$) slots earlier than user $B$. The blind rendezvous problem studied in this paper can be formulated as follows [9]:

$$\min \ t \ (t \in \mathbf{Z}^+), \tag{1a}$$

$$s.t. \ S_A^{(t+\delta-1) \bmod T_A + 1} = g, \tag{1b}$$

$$S_B^{(t-1) \bmod T_B + 1} = g, \tag{1c}$$

where $g \in (C_A \cap C_B)$ and $t$ means TTR which is the time span from the moment when both users start their channel hopping to the time when they achieve rendezvous for the first time.

## IV. QCMS-CH ALGORITHM

### A. Basic Idea

The CH sequence generated by the proposed QCMS-CH algorithm can be represented in a matrix structure, and we refer to this matrix as CH matrix. QCMS-CH takes two major steps to build the CH matrix as follows.

Step 1 is to generate a bootstrapping sequence. Each user first randomly selects an available channel and transforms its index (denoted by $R$) into a fixed-length quaternary number, then uses a specially constructed quaternary-quinary coding table to convert the quaternary number into a quinary number, and then the user adds the prefix "R00" in front of the quinary number to generate a bootstrapping sequence. The length of this bootstrapping sequence determines the number of columns of the CH matrix, and each element of the bootstrapping sequence, which may be one of {R, 0, 1, 2, 3, 4}, corresponds to a column of the CH matrix.

Step 2 is to build the CH matrix column by column. Since there are at most six different types of columns, i.e., R-type and λ-type (λ ∈ {0, 1, ..., 4}), six types of subsequences associated with R-type and λ-type can be elaborately generated based on the available channels of the user. Then, we can build the CH matrix by repeatedly placing the corresponding subsequence in each column. Once obtain the CH matrix, the user can access the matrix row by row and accordingly perform its channel hopping to attempt rendezvous with other users. That is, the final CH sequence can be obtained by concatenation each row of the CH matrix.

An illustration on the basic idea of QCMS-CH is given in Fig. 1. Next, we will detail the quaternary-quinary coding table, the generation of bootstrapping sequence as well as the CH matrix.

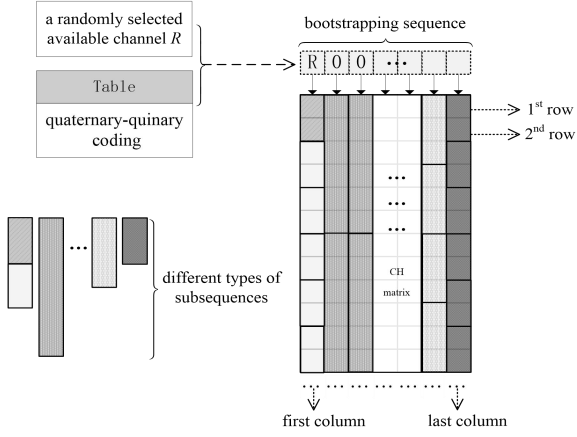

Fig. 1. Illustration of basic idea of QCMS-CH. In the CH matrix, each column repeats the corresponding subsequence. A little exceptional case is the 1st column that is associated with R-type subsequence, where the repeated items is not the whole but a half of the R-type subsequence.

*B. Quaternary-Quinary Coding*

Table I presents the quaternary-quinary coding, which will be used to generate the bootstrapping sequence. As shown in Table I, each 2-digit quaternary number can be transferred into a 2-digit quinary number. Since there are 16 (=$4^2$) different 2-digit quaternary numbers in total, the size of the quaternary-quinary coding is 16. Our coding is a bijection from 2-digit quaternary numbers to 2-digit quinary numbers, since the selected 2-digit quinary numbers should satisfy the following two conditions and have exact 16 different choices: 1) The first digit is not 0; 2) The first digit and the second digit are different.

Table I. Quaternary-quinary coding

| quaternary | quinary | quaternary | quinary |
|---|---|---|---|
| 00 | 10 | 20 | 30 |
| 01 | 12 | 21 | 31 |
| 02 | 13 | 22 | 32 |
| 03 | 14 | 23 | 34 |
| 10 | 20 | 30 | 40 |
| 11 | 21 | 31 | 41 |
| 12 | 23 | 32 | 42 |
| 13 | 24 | 33 | 43 |

The quaternary-quinary coding has two properties, which are given as follows. Due to limited space, we omit the proof of the properties. We give the illustrative examples of these two properties in Fig. 2.

**Property 1**. Given a sequence which is composed of multiple 2-digit quinary numbers in the quaternary-quinary coding (Table I), there are no three consecutive identical numbers in the sequence.

**Property 2**. Given a sequence which is composed of multiple 2-digit quinary numbers in the quaternary-quinary coding (Table I), there may be two consecutive identical numbers in this sequence. If so, the first number of the two consecutive identical numbers must be at an even position in the sequence.

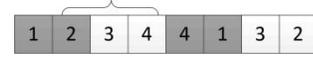

(a) No three consecutive identical numbers (for Property 1).

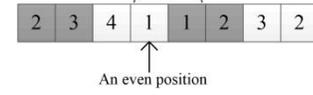

(b) Existing two consecutive identical numbers with the first number at an even position (for Property 2).

Fig. 2. Illustrative examples of Property 1 and Property 2.

*C. Generation of Bootstrapping Sequence*

Based on the aforementioned quaternary-quinary coding, we can describe the generation of bootstrapping sequence as follows (see Algorithm 1 for its pseudocode). Suppose a user randomly selects an available channel (index) $R$. As shown in Lines 1~3 of Algorithm 1, the decimal number $R$ should first be represented as a quaternary number $q$. The length of $q$ is an even number which is either $\lceil \log_4 N \rceil$ or $\lceil \log_4 N \rceil + 1$. Next, we may view $q$ as a string and group $q$ into a series of 2-digit quaternary numbers from left to right. Then, each 2-digit quaternary number in $q$ will be transferred into the corresponding 2-digit quinary number, and thereby we can obtain a quinary number (string) $Q$ (see Line 4 of Algorithm 1). Finally, we add a prefix "R00" in front of $Q$ and get the bootstrapping sequence (see Line 5 of Algorithm 1).

---

**Algorithm 1**: Generation of bootstrapping sequence

---

**Input**: the number of channels $N$, a randomly selected available channel $R$, and the quaternary-quinary coding (i.e., Table I).

**Output**: A bootstrapping sequence $BS$.

1: Represent $R$ as quaternary number $q$ of length $L_q = \lceil \log_4 N \rceil$;

2: **if** ($L_q$ is an odd number)

3:     Add a '0' in front of the quaternary number $q$;

4: Use Table I to convert $q$ into a quinary number $Q$ by taking two consecutive numbers of $q$ as a group;

5: $BS$ = "R00" || $Q$;  // "||"denotes the concatenation operation

6: **return** $BS$;

---

An example is given here to illustrate how the bootstrapping sequence is generated. We assume that the number of channels $N$=200 and the set of available channels of user $A$ is $C_A$={1, 2, 3, 4, 5, 6}. If user $A$ randomly selects an available channel $R$=5, $R$ will be represented as quaternary number $q$=**00**11 with length of $\lceil \log_4 200 \rceil$=4. Base on the quaternary-quinary coding in Table I, $q$ can be transferred into quinary number $Q$=**10**21 (**00** and 11 in $q$ correspond to **10** and 21 in $Q$, respectively). Then, adding a prefix "R00" in front of $Q$ will produce the bootstrapping sequence $BS$="R00**10**21"=(R, 0, 0, 1, 0, 2, 1).

Given a bootstrapping sequence $BS=(BS(1), BS(2), ..., BS(L))$ and an integer $d \in [1, L-1]$, we define its rotation as $Rotate(BS,d)=(BS(i_1), BS(i_2), ..., BS(i_L))$, where $i_j=(j-1+d)$ **mod** $L +1$ ($j=1, 2, ..., L$) and the mod denotes the modular operation. Then, the bootstrapping sequences generated by Algorithm 1 have the following property. An illustrative example of the property is given in Fig. 3.

**Property 3**. Given two bootstrapping sequences $BS_1$ and $BS_2$ that are generated by Algorithm 1 ($BS_1$ and $BS_2$ may be the same), let $S=Rotate(BS_2,d)$ for any $d \in [1, L-1]$, then there exist $i$ and $j$ such that: 1) $BS_1(i)=0$ and $S(i) \in \{1, 2, 3, 4\}$; 2) $S(j)=0$ and $BS_1(j) \in \{1, 2, 3, 4\}$. That is, there exist at least one overlap of number 0 and other number in $\{1, 2, 3, 4\}$ between $BS_1$ and $Rotate(BS_2,d)$.

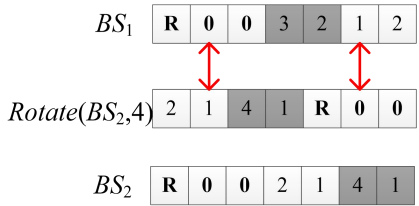

Fig. 3. An illustrative example of Property 3.

### D. Generation of Subsequences

There are six different types of subsequences which include R-type and λ-type ($\lambda \in \{0, 1, ..., 4\}$). Algorithm 2 and Algorithm 3 present the pseudocodes of the generation of R-type subsequence and λ-type subsequence, respectively.

According to Algorithm 2, it is clear that a R-type subsequence has a fix length of 10. Given a randomly selected available channel $R$, the first five items and the last item in the generated R-type subsequence are equal to $R$, and the others in the subsequence are denoted by a wildcard character '*'. Here, the wildcard character '*' means that it can be replaced by any available channel.

The λ-type subsequence is generated based on the available channels. Specifically, given a set of available channels $C \subseteq \mathbf{C}$, we should first select the smallest prime number $P \geq \max\{n=|C|, 5\}$ (see Line 1 in Algorithm 3), and then determine the length $K$ of the λ-type subsequence according to both $\lambda$ and $P$ (see Lines 2-7 in Algorithm 3). The first $n$ items in the the λ-type subsequence are a permutation of $C$ (see Lines 8 and 10 in Algorithm 3), and the others are denoted by wildcard character '*'.

For ease of understanding, we give the examples of R-type and λ-type subsequences in Fig. 4.

---

**Algorithm 2**: Generation of R-type subsequence

**Input**: an available channel $R$.
**Output**: A CH sequence $Seq=S(t)|_{t=1, 2, ..., 10}$.
1: **for** $t=1, 2, ..., 10$
2:    **if** ( $t \leq 5$ or $t=10$)     $S(t)=R$;
3:    **else**    $S(t)=$'*'; //wildcard character
4: **return** $Seq$;

---

**Algorithm 3**: Generation of λ-type subsequence

**Input**: an available channel set $C=\{C(1), C(2), ..., C(n)\} \subseteq \mathbf{C}$ and $\lambda \in \{0, 1, 2, 3, 4\}$.
**Output**: A CH sequence $Seq=S(t)|_{t=1, 2, ..., K}$.
1: Let $P$ is the smallest prime number $\geq \max\{n, 5\}$;
2: **switch** ($\lambda$) :
3:    **case** 0 : $K=P$;
4:    **case** 1 : $K=P+2$;
5:    **case** 2 : $K=P+3$;
6:    **case** 3 : $K=P+4$;
7:    **case** 4 : $K=P+6$;
8: $A = rearrange(C)$; //a permutation of $C$
9: **for** $t=1, 2, ..., K$
10:    **if** $t \leq n$    $S(t)=A(t)$;
11:    **else**    $S(t)=$'*'; //wildcard character
12: **return** $Seq$;

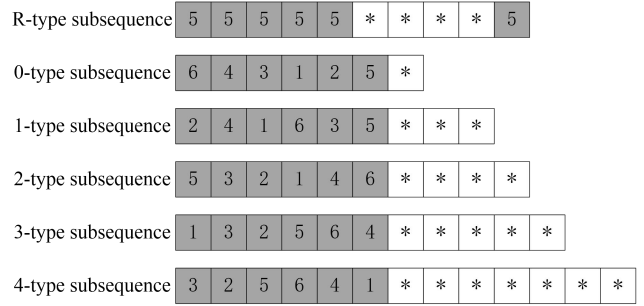

Fig. 4. Examples of R-type and λ-type subsequences ($\lambda=0, 1, ..., 4$) for user $A$ with available channel set $C_A=\{1, 2, 3, 4, 5, 6\}$ and a randomly selected available channel $R=5$. In each λ-type subsequence, the items in gray shadow consist of a permutation of $C_A$.

Regarding the R-type subsequences generated by Algorithm 2 and the λ-type subsequences generated by Algorithm 3, we have the following properties.

**Property 4**. Given a R-type subsequence $Seq$, we set $S=Seq(1:5)+Seq(6:10)+Seq(6:10)+...+Seq(6:10)+...$, where '+' denotes the sequence concatenation operator, then two users can achieve rendezvous within five slots by channel hopping according to sequence $S$.

An illustrative example of Property 4 is given in Fig. 5.

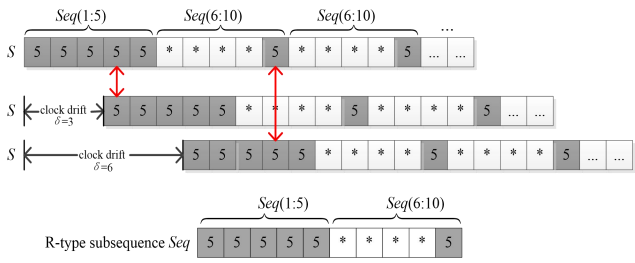

Fig. 5. An illustrative example of Property 4.

**Property 5**. If $P$ is a prime number not less than 5, then the five numbers $P$, $P+2$, $P+3$, $P+4$, $P+6$ are mutually coprime.

For example, suppose $P=7$, it can be easily verified that $P=7$, $P+2=9$, $P+3=10$, $P+4=11$, and $P+6=13$ are mutually coprime.

**Property 6**. Let $Seq_A$ and $Seq_B$ be two λ-type subsequences that users $A$ and $B$ generate based on their available channel sets $C_A$ and $C_B$, respectively, by using Algorithm 3. If the length of $Seq_A$ is coprime with that of $Seq_B$, then the two users can visit channel pair $x$-$y$ for any $x \in C_A$ and $y \in C_B$ within $|Seq_A| \times |Seq_B|$ slots when they implement $Seq_A$ and $Seq_B$ respectively for channel hopping.

An illustrative example of Property 6 is given in Fig. 6. In this example, we assume users $A$ and $B$ have the available channel sets $C_A=\{1, 2, 3, 4, 5, 6\}$ and $C_B=\{1, 7, 8, 9\}$, respectively. $Seq_A$ is a 1-type subsequence of user $A$, and $Seq_B$ is a 0-type subsequence of user $B$. We can see that $|Seq_A|=9$ and $|Seq_B|=5$. When the two users repeatedly implement $Seq_A$ and $Seq_B$ respectively, they can visit 24 different channel pairs $x$-$y$ for $x \in C_A$ and $y \in C_B$ within $|Seq_A| \times |Seq_B|=45$ slots. In particular, we can find that the two users rendezvous on channel 1 after 19 slots (see the line in red color), i.e., with TTR=19.

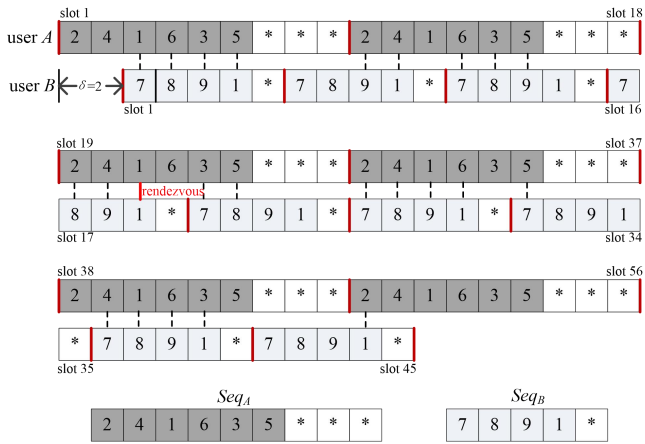

Fig. 6. An illustrative example of Property 6. Assume $C_A=\{1, 2, 3, 4, 5, 6\}$ and $C_B=\{1, 7, 8, 9\}$. $Seq_A$ is a 1-type subsequence of user $A$, and $Seq_B$ is a 0-type subsequence of user $B$.

### E. QCMS-CH

Based on Algorithms 1~3, i.e., the generation of bootstrapping sequence, R-type and λ-type subsequences, now we can present the QCMS-CH algorithm, with its pseudocode described in Algorithm 4. When a user implements the QCMS-CH algorithm, the user should first generate its bootstrapping sequence (see Line 2 in Algorithm 4), R-type and λ-type subsequences (see Lines 4-6 in Algorithm 4). Then, the user will perform its channel hopping based on these subsequences to attempt rendezvous (see Lines 8-20 in Algorithm 4).

As pointed out in our basic idea (Section IV.A), a user that implements the QCMS-CH algorithm actually performs its channel hopping by periodically traversing the associated CH matrix (denoted by $CHM$) in a row-by-row manner. The number of columns of $CHM$ equals $L$, i.e., the length of the bootstrapping sequence. Except the 1st column (associated with R-type subsequence), the other $i$th column of $CHM$ is a concatenation of copies of subsequence $Seq^{(i)}$, which can be expressed as $CHM(:, i)=Seq^{(i)}+Seq^{(i)}+...+Seq^{(i)}+...$ ($i=2, 3, ..., L$), where '+' denotes the sequence concatenation operator. For the 1st column, which is associated with R-type subsequence, it can be expressed as $CHM(:, 1)=Seq^{(1)}(1:5)+Seq^{(1)}(6:10)+...+Seq^{(1)}(6:10)+...$, that is, the repeated items is not the whole but a half of the R-type subsequence $Seq^{(1)}$.

| **Algorithm 4**: QCMS-CH |
|---|
| **Input**: the number of channels $N$, an available channel set $\mathsf{C}=\{C(1), C(2), ..., C(n)\} \subseteq \boldsymbol{C}$, the quaternary-quinary coding (i.e., Table I) |
| 1: Randomly select a channel $R$ from $\mathsf{C}$; |
| 2: $BS$=**Algorithm 1**($N$, $R$, Table I); //bootstrapping sequence |
| 3: $L$=length of $BS$; |
| 4: **for** $i$=1, 2, ..., $L$   //generate L subsequences |
| 5:   **if** ($BS(i)$='R')   $Seq^{(i)}$=**Algorithm 2**($R$); //R-type |
| 6:   **else**   λ=$BS(i)$; $Seq^{(i)}$=**Algorithm 3**($\mathsf{C}$, λ); //λ-type |
| 7: $t$=0; |
| 8: **while** (not rendezvous)   //channel hopping |
| 9:   $i$=($t$ **mod** $L$)+1;   //in the ith column of CH matrix |
| 10:   $j$=⌈$t/L$⌉+1;   //in the jth row of CH matrix |
| 11:   **if** $BS(i)$='R'   // implement R-type subsequence |
| 12:     **if** $j \leq 5$ |
| 13:       $c$=$Seq^{(i)}(j)$; |
| 14:     **else** |
| 15:       $k$=($j$−6) **mod** 5+6; $c$=$Seq^{(i)}(k)$; |
| 16:   **else**   //implement λ-type subsequence |
| 17:     $k$=($j$−1) **mod** $|Seq^{(i)}|$+1; $c$=$Seq^{(i)}(k)$; |
| 18:   **if** ($c$='*')   //wildcard character |
| 19:     $c$=a channel randomly selected from $\mathsf{C}$; |
| 20:   Attempt rendezvous on channel $c$; |

Now, let's give some examples to show how the proposed QCMS-CH algorithm works. Again, we assume users $A$ and $B$ have available channel sets $C_A = \{1, 2, 3, 4, 5, 6\}$ and $C_B = \{1, 7, 8, 9\}$, respectively. Suppose users $A$ and $B$ randomly select available channel $R_A=5$ and $R_B=1$, respectively. Fig. 7(a) shows the bootstrapping sequences and the CH matrices of users $A$ and $B$, and Fig. 7(b) shows the channel hopping sequences of the two users.

Based on the aforementioned Properties 1-6, we can derive an upper-bound of maximum TTR of the QCMS-CH algorithm, which is presented in Theorem 1 as follows.

**Theorem 1.** If users $A$ and $B$ implement the QCMS-CH algorithm (i.e. Algorithm 4) based on their available channel sets $C_A$ and $C_B$, respectively, they can achieve rendezvous within $\max\{(P_A+4)(P_B+6), (P_A+6)(P_B+4)\} \times L$ slots, where $L = 2 \times \lceil \lceil \log_4 N \rceil / 2 \rceil + 3$, $P_A$ and $P_B$ are the smallest prime numbers which are not less than $\max\{5, |C_A|\}$ and $\max\{5, |C_B|\}$, respectively.

Based on the setting of Fig. 7, we present several examples in Fig. 8 to further show that QCMS-CH can guarantee rendezvous with short TTR. Notice that users $A$ and $B$ have only one common channel (i.e., channel 1) in this case, which is a difficult situation for the two users to achieve rendezvous. According to Theorem 1, the maximum TTR can be upper bounded by $\max\{(P_A+4)(P_B+6), (P_A+6)(P_B+4)\} \times L$ which is equal to 847 in this case. As shown in Fig. 8, by implementing QCMS-CH users $A$ and $B$ can achieve rendezvous in various scenarios. For example, the two users achieve rendezvous with TTR=15 when clock drift is 2 slots (see Fig. 8(b)) and achieve rendezvous with TTR=22 when clock drift is 5 slots (see Fig. 8(c)). Here, we can find that the actual TTRs are far less than the upper bound of MTTR, which means that the upper bound of MTTR derived in Theorem 1 is a loose bound. We will further demonstrate the excellent performance of QCMS-CH later through extensive simulations.

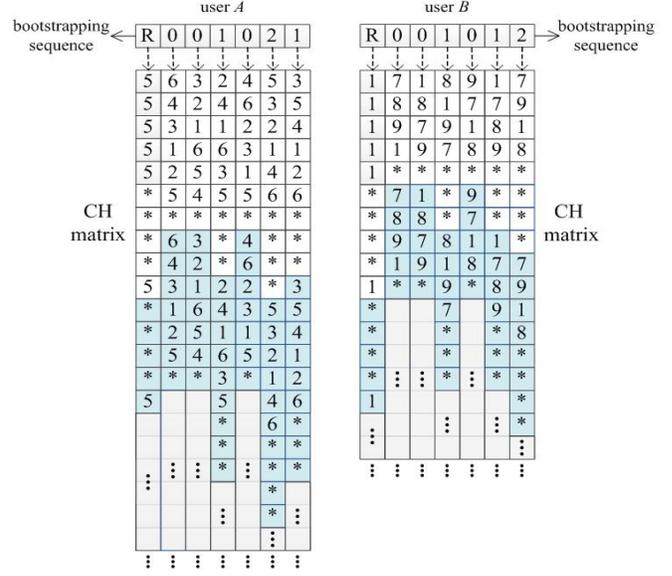

(a) Bootstrapping sequences and CH matrices of two users.

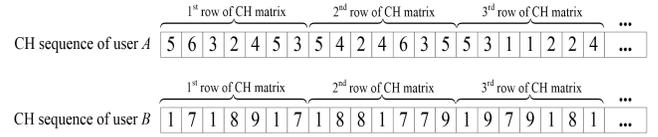

(b) CH sequences that two users implements.

Fig. 7. An illustrative example that users $A$ and $B$ implement the QCMS-CH algorithm (i.e., Algorithm 4). Assume $C_A=\{1, 2, 3, 4, 5, 6\}$ and $C_B=\{1, 7, 8, 9\}$; users $A$ and $B$ randomly select available channel $R_A=5$ and $R_B=1$, respectively.

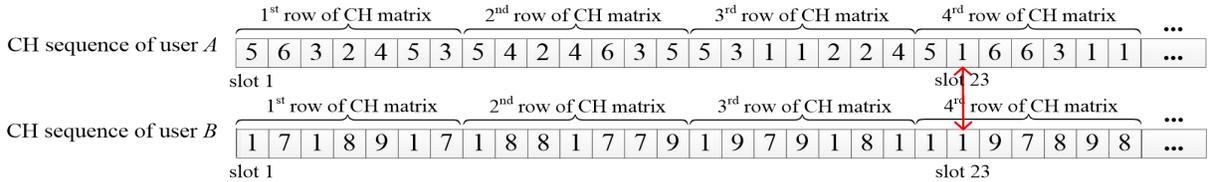

(a) clock drift $\delta=0$, users $A$ and $B$ achieve rendezvous with TTR=23.

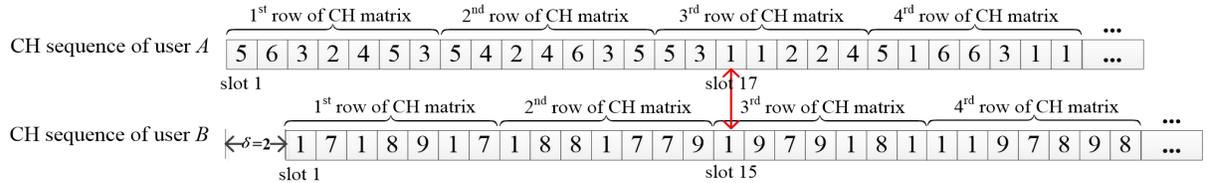

(b) clock drift $\delta=2$, users $A$ and $B$ achieve rendezvous with TTR=15.

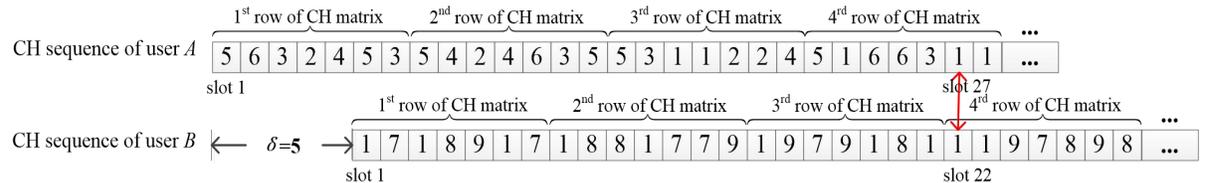

(c) clock drift $\delta=5$, users $A$ and $B$ achieve rendezvous with TTR=22.

Fig. 8. An illustrative example of guaranteed rendezvous of QCMS-CH.

Table II. Comparisons of different algorithms on MTTR upper-bound.

| CH algorithm | Common features | MTTR upper-bound |
|---|---|---|
| ABIO [6] | ✓ symmetric role<br>✓ heterogeneous channels<br>✓ asynchronous clock<br>✓ anonymous ID<br>✓ single-radio<br>✓ local channel set | MTTR$\leq$max$\{P_A Q_B, P_B Q_A\} \times (3L+1)$.<br>$L = \lfloor \log_2 N \rfloor + 1$. |
| EEA [5] | | MTTR$\leq M \times P_{A,1} \times P_{B,1}$.<br>$M = 2 \times \lceil \log_2 \lceil \log_2 N \rceil \rceil + 7$. |
| QR [7] | | MTTR$\leq M \times P_{A,1} \times P_{B,1}$.<br>$M = 5 \times \lceil \lceil \log_2 N \rceil/4 \rceil + 6$. |
| IQSF-CH [8] | | MTTR$\leq F(P_i, P_j) \times (2 \times \lceil \log_2 N \rceil + 3)$.<br>$F(P_i,P_j) = \begin{cases} \max\{(n_j-G)P_i + 2P_i - 1, (n_i P_i - GP + 1)P_j\} & \text{if } P_i < P_j \\ (\max\{n_i, n_j\} - G + 1)P_j & \text{if } P_i = P_j \\ \max\{(n_i-G)P_j + 2P_j - 1, (n_j P_j - GP_j + 1)P_i\} & \text{if } P_i > P_j \end{cases}$ |
| QECH [9] | | MTTR$\leq$max$\{L_A, L_B\} \times$max$\{P_A(P_B+4), P_B(P_A+4)\}$.<br>$L_A = 2^{\lceil \log_2(4+4\lceil \log_2 R_A \rceil/6\rceil) \rceil}$.<br>$L_B = 2^{\lceil \log_2(4+4\lceil \log_2 R_B \rceil/6\rceil) \rceil}$. |
| **QCMS-CH** | | **MTTR$\leq$max$\{(P_A+4)(P_B+6),(P_B+4)(P_A+6)\} \times L$.<br>$L = 2 \times \lceil \lceil \log_4 N \rceil/2 \rceil + 3$.** |

**Remarks**: $N$ is the number of global available channels; $G$ is the number of common available channels of two users; $Q_A$ and $Q_B$ are the smallest integer not less than the number of local available channels of user $A$ and user $B$, respectively, and the factor of $Q_A$ and $Q_B$ only contain 2 or 3; $n_i$ and $n_j$ are the number of local available channels of user $i$ and user $j$, respectively; $P_i$ and $P_j$ are the smallest prime number not less than $n_i$ and $n_j$, respectively; $P_A$ and $P_B$ are the smallest prime number not less than the number of local available channels of user $A$ and user $B$, respectively, in addition, $P_A$ and $P_B$ must not be less than 5. $P_{A,1}$ and $P_{B,1}$ are the second smallest prime number among the prime numbers that is not less than the number of local available channels of user $A$ and user $B$, respectively; $R_A$ and $R_B$ are the value corresponding to the channel randomly selected by user $A$ and user $B$, respectively.

## V. PERFORMANCE EVALUATION

In this section, the proposed QCMS-CH algorithm is compared with five state-of-the-art rendezvous algorithms, i.e., EEA [5], ABIO [6], QR [7], IQSF-CH [8], and QECH [9]. Table II summarizes the related features and MTTR upper bounds of these algorithms.

Recall that $N$ is the number of all potentially available channels, $C_A$ and $C_B$ are the sets of available channels of user $A$ and user $B$, respectively. We introduce $\theta_A = |C_A|/N$ and $\theta_B = |C_B|/N$ to denote the available channel ratios of users $A$ and user $B$, respectively. In the simulation, we will vary both $\theta_A$ and $\theta_B$ to generate different sets of available channels for users $A$ and $B$.

We use MATLAB 9.8.0 (R2020a) to conduct the simulation. We mainly consider three different scenes with various settings of parameters. Each algorithm is independently executed 30000 times for every setting of parameters, and in each run the clock drift between the two users is randomly selected from [0, 50] so as to simulate the asynchronous clock.

### A. Scene 1: Influence of $G_{AB}$ for Fixed $\theta_A$, $\theta_B$ and N

In Scene 1, we fix $\theta_A$, $\theta_B$ and $N$, and observe the influence of varying $G_{AB}$ on MTTR and ETTR. We set the parameters $\theta_A=0.3$, $\theta_B=0.4$, $N=200$ and $G_{AB} \in [1,10]$. Fig. 9(a) compares the MTTRs of different algorithms. When $G_{AB}$ increases from 1 to 4, the MTTRs of almost all algorithms gradually decrease, and the MTTR of QCMS-CH is the smallest among the six algorithms. In particular, when $G_{AB}=1$, i.e., there is only one channel commonly available to the two users, which is a very tough case for the users to achieve rendezvous, we can find that our QCMS-CH algorithm significantly outperforms the others.

Fig. 9(b) compares the ETTRs of different algorithms. As $G_{AB}$ increases, the ETTRs of all algorithms are gradually decreasing. We can see that the ETTR of QCMS-CH is always the smallest. For example, when $G_{AB}=1$, the ETTR of the second best algorithm QECH (=4722) is 4.31% larger than that of QCMS-CH (=4527). When $G_{AB}=5$, the ETTR of QECH (=1013) is 5.30% larger than that of QCMS-CH (=962). When $G_{AB}=10$, the ETTR of QECH (=513) is 7.55% larger than that of QCMS-CH (=477). It can be seen from the above observations that the ETTRs of other algorithms are on average 5.72% larger than that of QCMS-CH, and the ETTR performance advantage of QCMS-CH is hardly affected by the change of $G_{AB}$.

### B. Scene 2: Influence of N for Fixed $\theta_A$, $\theta_B$ and $G_{AB}$

In Scene 2, we fix $\theta_A$, $\theta_B$ and $G_{AB}$, and observe the influence of varying $N$ on MTTR and ETTR. We set the parameters $\theta_A=0.3$, $\theta_B=0.4$, and $N \in [40, 220]$. Fig. 10(a) compares the MTTRs of different algorithms. We can see the MTTRs of all algorithm increase as $N$ increases and QCMS-CH has the best MTTR. The algorithm QECH is based on local available channel sets rather than global available channel sets to reduce the MTTR. However, in fact, its bootstrapping sequence length needs to be padded to a power of two, so its bootstrapping sequence is longer sometimes, which may result in relatively large MTTR.

Fig. 10(b) compares the ETTRs of different algorithms. As $N$ increases, the ETTRs of all algorithms also increase. In all cases, the ETTR of QCMS-CH is the smallest, which clearly shows the superiority of QCMS-CH. The theoretical MTTR upper-bound of EEA is similar to QCMS-CH in this case (see Table II), but EEA will repeatedly select two channels in multiple time slots [5]. Thus, compared with QCMS-CH, EEA may have a lot of redundancy in its CH sequence, resulting in much larger ETTR.

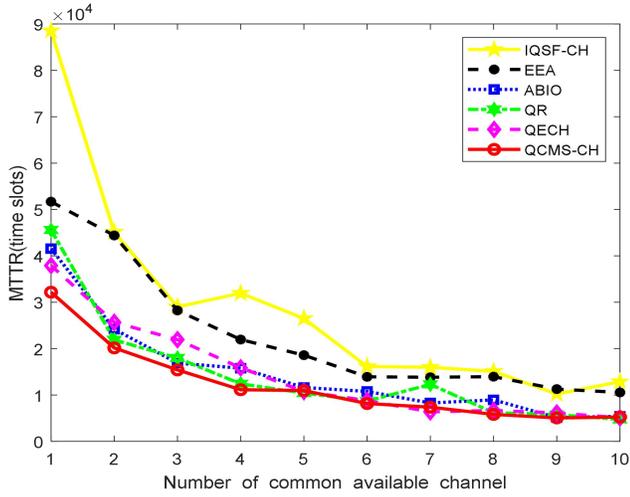
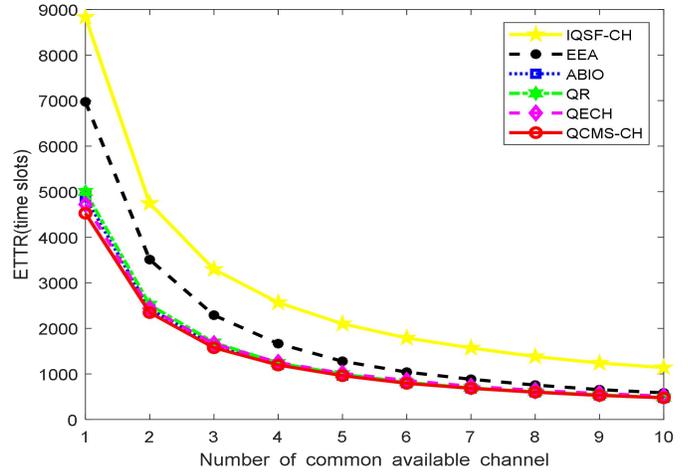

(a) MTTR VS. $G_{AB}$

(b) ETTR VS. $G_{AB}$

Fig. 9. The influence of $G_{AB}$

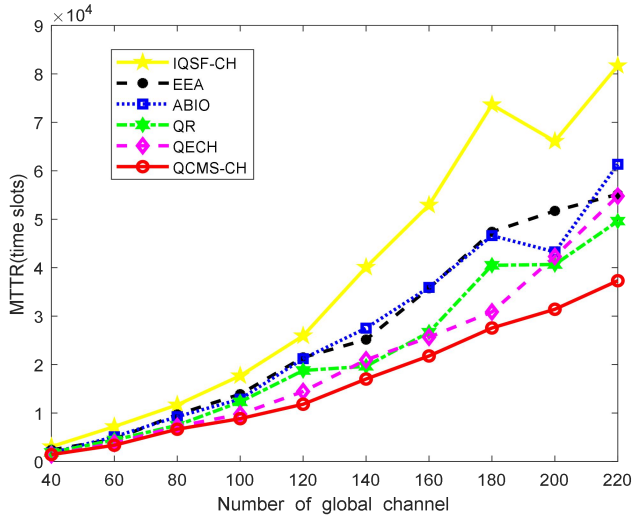
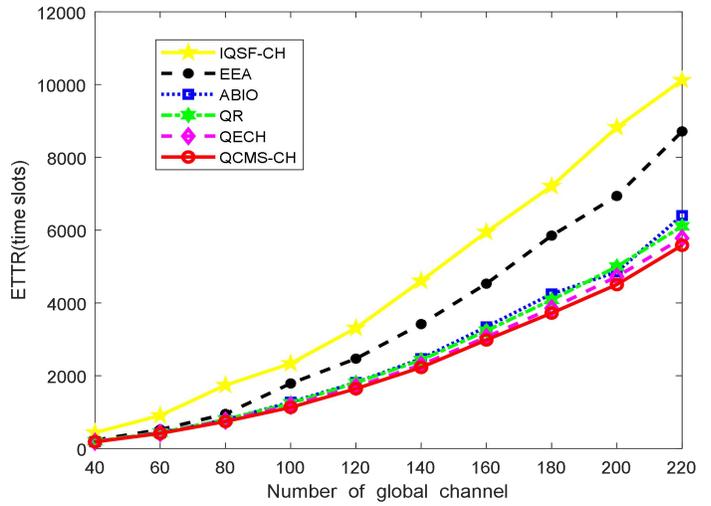

(a) MTTR VS. $N$

(b) ETTR VS. $N$

Fig. 10. The influence of $N$

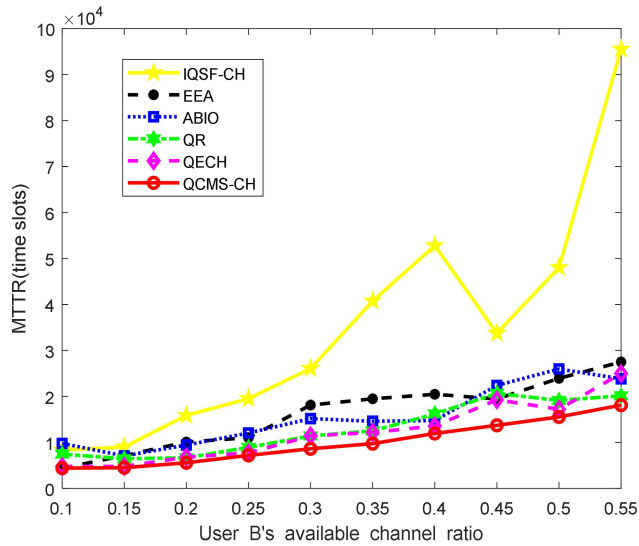
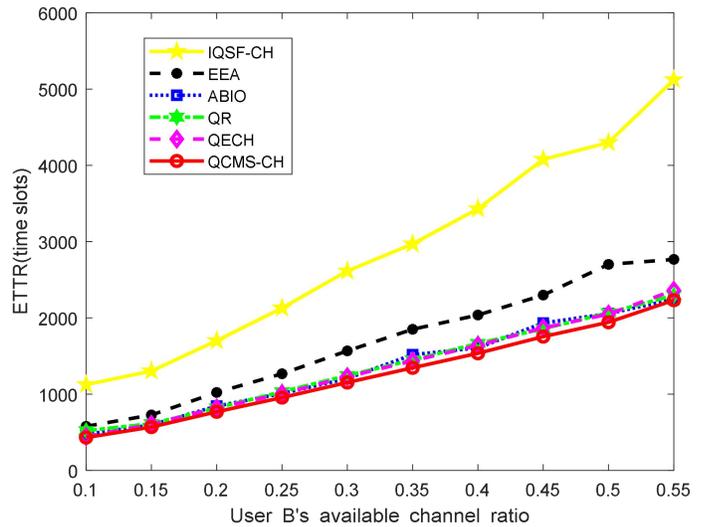

(a) MTTR VS. $\theta_B$

(b) ETTR VS. $\theta_B$

Fig. 11. The influence of $\theta_B$

## C. Scene 3: Influence of $\theta_B$ for Fixed $\theta_A$, N and $G_{AB}$

In Scene 3, we fix $\theta_A$, $N$ and $G_{AB}$, and observe the influence of varying $\theta_B$ on MTTR and ETTR. We set the parameters $\theta_A$=0.1, $\theta_B \in [0.1, 0.55]$, $N$=200, and $G_{AB}$=1. Fig. 11(a) compares the MTTRs of different algorithms. In most cases, the MTTRs of all algorithms increase with the increase of $\theta_B$. The MTTR of QCMS-CH is still always the smallest, which shows the stability of the performance advantage.

Fig. 11(b) compares the ETTRs of different algorithms. The ETTRs of all algorithms almost increase proportionally with the increase of $\theta_B$. The ETTR of QCMS-CH is always the smallest among the six algorithms. Specifically, when $\theta_B$=0.1, the ETTR of QECH (=444) is 3.50% larger than that of QCMS-CH (=429). When $\theta_B$=0.3, the ETTR of QECH (=1228) is 6.50% larger than that of QCMS-CH (=1153). When $\theta_B$=0.5, the ETTR of QECH (=2049) is 5.46% larger than that of QCMS-CH (=1943). In this scene, the ETTR of QCMS-CH is on average 5.15% better than the compared algorithms.

## VI. CONCLUSION

In this paper, we propose a novel quinary coding and matrix structure-based CH algorithm (QCMS-CH) for blind rendezvous in DCRNs. QCMS-CH is applicable to single-radio at each user, asynchronous clocks, heterogeneous available channels, and symmetric roles among the users. Moreover, QCMS-CH depends on only the available channels of user and does not need using user ID. Our theoretical analysis show that QCMS-CH can guarantee rendezvous with relatively small MTTR upper bound. Extensive simulation results have shown the superiority of QCMS-CH in terms of both MTTR and ETTR. In particular, compared with the state-of-the-art, QCMS-CH can achieve 5% reduction in terms of ETTR in many cases.

## ACKNOWLEDGMENT

This research was supported by Key Projects of Colleges and Universities in Guangdong (No. 2019KZDXM063), the National Natural Science Foundation of China (No. 61872096), Guangzhou Science and Technology Plan Project (No. 202102080348), 2021 China International College Students' Innovation and Entrepreneurship training program, and 2022 Special fund support project for scientific and technological innovation cultivation of Guangdong University Students (No. pdjh2022b0298).